\documentstyle[twoside,fleqn,espcrc2]{article}
\input epsf
\title{String Physics and Black Holes}

\author{Leonard Susskind and John Uglum\address{Department of
Physics, Stanford University \\
Stanford, CA  94305-4060  USA}}

\def\av{\vec \alpha}
\def\kv{\vec k}
\def\pa{\partial}

\def\Pv{\vec P}
\def\tr{\hbox{tr}}
\def\xv{\vec x}
\def\Xv{\vec X}

\begin{document}

\begin{abstract}

In these lectures we review the quantum physics of large
Schwarzschild black holes.  Hawking's information paradox, the theory
of the stretched horizon and the principle of black hole
complementarity are covered.  We then discuss how the ideas of black
hole complementarity may be realized in string theory.  Finally,
arguments are given that the world may be a hologram.

\end{abstract}

\maketitle

\section{Introduction}

An outsider listening to this conference might get the idea that
there is such a thing as string theory, that string theory is a
relatively complete theory of the world, including gravitation.  We
do not believe this is so, at least not yet.  There is a wide class
of phenomena, perhaps the most interesting phenomena for future
study, which string theory in its present formulation cannot address
at this time, and perhaps cannot address at all--Planck scale
physics.

The distinction between Planck scale physics and string scale physics
is often ignored in string theory, but it is an imporant one.  If $g$
is the string coupling and the number of large (uncompactified)
spacetime dimensions is $D$, then the Planck scale $\ell_P$ is
related to the string scale $\ell_S$ by the relation $\ell_P^{D-2} =
g^2 \ell_S^{D-2}$.  It is usually assumed that the string coupling is
very small, so there is a large difference between the two scales.
Most of the phenomena that we are able to discuss in string
theory--the spectrum, scattering amplitudes, etc. are all phenomena
that have to do with the strings scale, not the Planck scale.

There is a host of problems at the Planck scale which the present
formulation of string theory is simply incapable of handling.  These
include

\begin{itemize}

\item{The thermodynamics of strings and their behavior at
temperatures above the Hagedorn temperature \cite{aw}}

\item{Very high energy scattering processes at very small impact
parameter \cite{gm}}

\item{Black hole evaporation and the puzzles associated with it}

\item{The cosmological constant problem}

\end{itemize}

In the case of a high energy central collision between two strings,
it is easy to guess the answer--a black hole forms, and then
evaporates.  But we can't study this in string theory.  If we did
understand quantum gravity, then we could answer all of these
questions.

The plan of these lectures is as follows.  First, we will review the
physics of horizons, including their thermal behavior, and all of the
fundamental physics of large mass Schwarzschild black holes.  We will
see that all of the paradoxes associated with black hole evaporation
can be addressed in this simple context, and attempts to resolve
these paradoxes will lead us to the idea of black hole
complementarity.

After reviewing the light front gauge formulation of strings, we will
be able to ask how string theory might be able to resolve the
paradoxes of black hole evaporation.  In particular, we will be
interested in how string theory stores information, and how the
Planck scale is generated from string theory.  We will discuss the
entropy of horizons in string theory.  Finally, we will discuss some
ideas about the world as a hologram, due to 't~Hooft and Susskind.

If one were to plot the distance scales that are probed as one
increases the energy of a process, we know that for ordinary
relativistic field theory, the length scale decreases as the energy
increases.  What we are now finding in string theory, however, is
that this behavior does not continue forever.  Increasing the energy
beyond the Planck energy, one starts to probe larger distances
instead of smaller ones.  This is the energy region which we must
understand to solve the aforementioned problems.

\section{Schwarzschild Black Holes}

The line element for the eternal Schwarzschild black hole geometry,
in Scwarzschild coordinates $(t,r, \theta, \varphi)$, is given by
\begin{eqnarray}\label{eqn:schmet}
ds^2 =& - \left ( 1 - \frac{2GM}{r}
\right ) dt^2 + \left ( 1 - \frac{2GM} {r} \right )^{-1}
dr^2\nonumber
\\ {} &+ r^2 d\Omega^2
\end{eqnarray}
where $G$ is Newton's constant, $M$ is the black hole mass, and
$d\Omega^2$ is the line element of the unit two-sphere.  The surface
$r = 2GM, t = \infty$ is the future event horizon.  There is also a
past event horizon at $r = 2GM, t = -\infty$, but we will not concern
ourselves much with this, since for black holes formed by the
gravitational collapse of matter it is absent.  The singularity is at
$r = 0$.

Light signals from points outside the horizon can reach infinity,
whereas light signals from points inside the horizon necessarily
terminate when they reach the singularity.  We can therefore think of
the horizon as consisting of those photons which were just barely
trapped by the black hole.  Physics is complicated near the
singularity, so we will restrict our attention to physics strictly
outside the black hole.  It is our belief that most of the
interesting physics is at the horizon, anyway.

To a freely falling observer, there is nothing special about the
horizon.  All of the local geometrical invariants remain small at the
horizon, so there is no local signal that he has crossed into the
region of the black hole, and he can cross the horizon into the black
hole in a finite amount of proper time.  Note, however, that $t =
\infty$ on the horizon, so an external observer, whose proper time is
proportional to $t$, will never see anything cross the horizon.  This
is the first of a series of peculiar situations in which observers
inside and outside will disagree.

Suppose we are interested in a region of space very close to the
horizon, small compared to the size of the black hole, but large
compared to any microscopic scales.  (Equivalently, we might be
interested in a black hole with very large $M$.)  In this case, the
Schwarzschild line element can be simplified.  If we define $\rho$ to
be the proper distance from the event
horizon,
\begin{eqnarray}
\rho &=& \sqrt{r (r - 2GM) } \nonumber \\ {} &{}& + 2GM \ln \left (
\frac{\sqrt{r - 2GM} + \sqrt{r} }{\sqrt{ 2GM} } \right )
\end{eqnarray}
and define a rescaled time coordinate $\omega = \frac{t}{4GM}$, then
for $r - 2GM \ll GM$, Eq.~(\ref{eqn:schmet}) can be approximated as
\begin{equation}\label{eqn:rindmet}
ds^2 = -\rho^2 d\omega^2 + d\rho^2 + dy^2 + dz^2 \>,
\end{equation}
where $y$ and $z$ denote the directions tangent to the horizon.
Eq.~(\ref{eqn:rindmet}) is known as the Rindler metric, and is
nothing more than flat Minkowski space in hyperbolic polar
coordinates.  If $(T, X, Y, Z)$ denote the coordinates of Minkowski
space, then
\begin{eqnarray}
T &=& \rho \sinh ( \omega ) \>,
\nonumber \\ {} X &=& \rho \cosh ( \omega ) \>, \nonumber \\ {} Y &=&
y
\>, \qquad Z = z \>.
\end{eqnarray}
A freely falling observer simply corresponds to an intertial
Minkowski observer, and there is clearly nothing special about the
horizon, which is just an ordinary light-like surface.

\begin{figure}
\vbox{{\centerline{\epsfsize=88cm \epsfbox{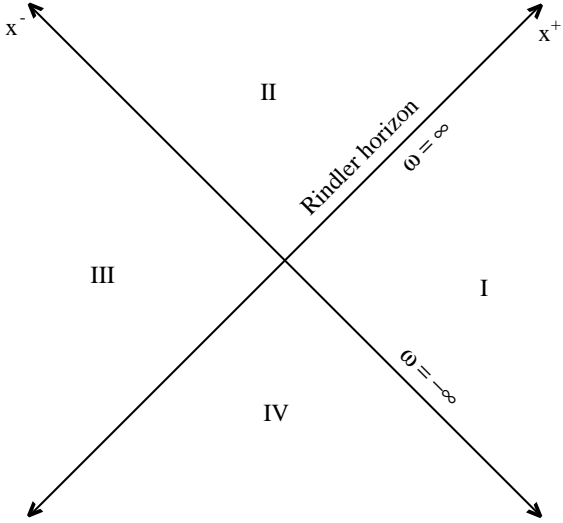}}}}
\caption{A diagram of Rindler space showing the four quadrants and
the horizon.}
\label{fig:rindlerspace}
\end{figure}

There is, however, something interesting about our parametrization of
flat space.  The time coordinate $\omega$ does not behave like an
ordinary Minkowski time variable--in fact, it corresponds to a
Lorentz boost parameter, and goes to infinity on the light like
surface of the horizon.  Spacelike surfaces of constant $\omega$
accumulate near the horizon.  Observers at fixed $\rho$ describe
hyperbolic trajectories, which means that they correspond to
uniformly accelerated observers in Minkowski space.  This makes
sense, of course, since an observer who wants to remain outside the
black hole must have a rocket or some other means of propulsion to
keep from falling in.  A diagram of Rindler space is shown in
Fig.~\ref{fig:rindlerspace}.

The classical physics of observers who are restricted to remain only
in the first quadrant of Rindler space, corresponding to the region
outside the black hole, is completely consistent.  Although signals
which originate in quadrant IV can influence events in quadrant I,
they must cross the past horizon at $\omega = -\infty$, and can
therefore be treated as initial data.  Similarly, signals which
propagate out from quadrant I into quadrant II must cross the surface
$\omega = \infty$.  Quadrant III is simply out of causal contact with
quadrant I, and signals originating there have no effect on events in
quadrant I.  When it comes to doing quantum mechanics in Rindler
space, however, the story will get more complicated.

\begin{figure}
\vbox{{\centerline{\epsfsize=88cm \epsfbox{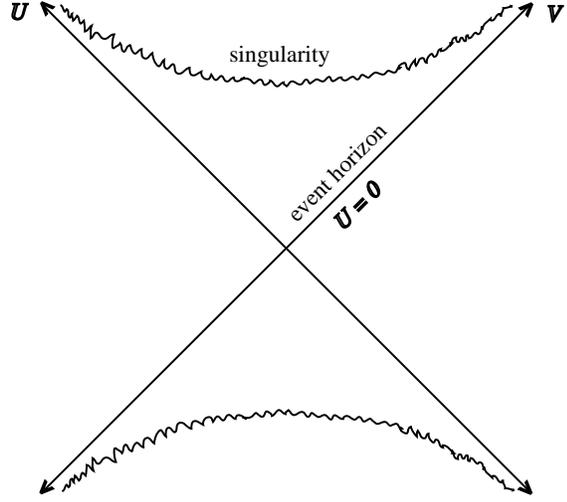}}}}
\caption{The Schwarzschild geometry in Kruskal coordinates.}
\label{fig:kruskal}
\end{figure}

Now that we have some new intuition about the nature of spacetime
near the horizon of a black hole, let's return to the full
Schwarzschild metric and see if we can make some of the ideas
obtained using the Rindler space approximation more precise.  Let us
first define the Regge-Wheeler tortoise coordinate
\begin{equation}\label{eqn:rwtort}
r_* = r + 2GM \ln ( r/2GM - 1 )
\end{equation}
and the Kruskal-Szekeres coordinates
\begin{eqnarray}\label{eqn:kzcoord}
U &=& - \exp \left ( (r_* - t)/4GM
\right ) \>, \nonumber \\ {}
V &=& \exp \left ( (r_* + t)/4GM \right )
\>.
\end{eqnarray}
Then the line element can be written
\begin{equation}
ds^2 = -\frac{32 G^2 M^3 e^{-r/2GM}}{r} dU dV + r^2 d\Omega^2 \>.
\end{equation}
The metric in the $(U, V)$ plane is conformal to flat space, and
nothing special happens at the horizon $r = 2GM$.  Note, however,
that the singularity at $r = 0$ has reappeared.
Fig.~\ref{fig:kruskal} shows what the geometry looks like in the $(U,
V)$ plane.

Using this diagram it is easy to understand the disagreement between
freely falling and external observers.  A freely falling observer
simply follows a geodesic, crossing the horizon in a finite amount of
proper time and eventually crashing into the singularity.  When the
freely falling observer crosses the horizon, he will no longer be
able to send signals to the outside world.  The external observer, in
order to avoid the black hole, must be constantly accelerating, and
therefore follows a hyperbolic trajectory.  The horizon corresponds
to a null surface which the external observer cannot intersect in a
finite time, and thus must be at $t = \infty$.  An external observer
can only receive signals from points outside the horizon, and can
thus never ``see'' anything cross the horizon.

In addition, as with the Rindler case, the surfaces of constant time
accumulate near the horizon, and the time coordinate $t$ actually
corresponds to a kind of Lorentz boost parameter.  As $t$ increases,
the relative boost between the external observer and an infalling
particle increases.  It is easy to show that as the particle falls
toward the horizon, its momentum as seen by the external observer
increases like $\exp ( t/4GM )$.  The black hole is the ultimate
particle accelerator--the momentum of any particle as seen by the
external observer will eventually become much larger than the Planck
mass.

\section{Quantum Physics in Rindler Space}

Now let us consider the question of how to do quantum mechanics in
Rindler space.  Consider a single real free massless scalar field
$\phi$ propagating in Rindler space.  It will prove convenient to
consider the wave equation for $\phi$ using the tortoise coordinates
$(t, r_*, \theta, \varphi)$.  The reason for this is that as $r_*$
runs from $-\infty$ to $\infty$, it covers the region from the
horizon out to infinity.  Thus it covers only the region outside the
black hole.  In these coordinates, the line element is
\begin{equation}
ds^2 = \left ( 1 - \frac{2GM}{r} \right ) [ -dt^2 + dr_*^2 ] + r^2
d\Omega^2 \>,
\end{equation}
and we see that the $(t, r_*)$ part of the metric is conformal to
flat space.  Thus, for a massless scalar field, we expect the wave
equation to take on a simple form.  If we define $\psi = r \phi$ and
expand $\psi$ in spherical harmonics $Y_{\ell}^m$, then the action
for $\psi_{\ell}^m$ is
\begin{eqnarray}
S =& \sum_{\ell} \frac{1}{2} \int dr_* dt \bigl [ ( \pa_t \psi_{\ell}
)^2 - ( \pa_{r_*}
\psi_{\ell})^2 \nonumber \\ {}
&- V_{\ell} (r_* ) ( \psi_{\ell} )^2 \bigr ]
\end{eqnarray}
where
\begin{equation}
V_{\ell} (r_*) = \frac{2GM}{r} \left [ \frac{\ell ( \ell + 1 ) }{r^2}
- \frac{2GM}{r^3} \right ] \>.
\end{equation}
The new feature is $V_{\ell}$, which has the form of a
position-dependent mass term for $\psi$.  $V_{\ell}$ is the
relativistic generalization of a centrifugal barrier, but it behaves
differently than an ordinary centrifugal barrier, because while it is
repulsive far from the black hole, it is {\sl attractive} for $r <
3GM$.  This means that particles of high angular momentum can be
trapped in the region between the horizon and $r = 3GM$, and can
rattle around in this region.

It is enlightening to examine the wave equation using the Rindler
approximation.  If we define a variable $u = \ln ( \rho )$, then $u$
behaves like $r_*$, in that it goes to $\infty$ at asymptotic
infinity and $-\infty$ at the horizon.  We can Fourier expand the
field $\phi$ to obtain transverse momentum modes $\phi_{\kv}$.  In
the coordinates $(\omega, u)$, the Lagrangian for $\phi_{\kv}$ takes
a particularly simple form,
\begin{equation}
{\cal L} = \frac{1}{2}\left [ (\pa_{\omega} \phi_{\kv})^2 - (\pa_u
\phi_{\kv})^2 - {\kv}^2 e^{2u} \phi_{\kv}^2 \right ] \>.
\end{equation}
The potential barrier is now given by $V_{\kv} = \kv^2 e^{2u}$, and
we see that only the mode with $\kv = 0$ can escape to infinity.

The next important piece of information we will need is the thermal
nature of Rindler space.  Specifically, we will see that a Rindler
observer describes the ordinary Minkowski vacuum by a thermal density
operator.  This is a general result, not restricted to the above case
of a massless scalar field.  Consider dividing the
hypersurface $T = 0$ of Minkowski space into two halves, one with $X
< 0$, and one with $X > 0$.  We will call these halves the left and
right halves, respectively.  Assume that the Hilbert space ${\cal H}$
on the hypersurface $T = 0$ factorizes into a product space ${\cal
H}_L \otimes {\cal H}_R$.  If $\{ | b \rangle_L \}$ is an orthonormal
basis for ${\cal H}_L$ and $\{ | a \rangle_R \}$ is an orthonormal
basis for ${\cal H}_R$, then a general ket $| \psi \rangle$ in ${\cal
H}$ can be written
\begin{equation}
| \psi \rangle = \sum_{b, a} \psi (b, a) | b \rangle_L \otimes | a
\rangle_R \>.
\end{equation}
If we now trace over the degrees of freedom in ${\cal H}_L$, the
resulting density matrix for the right half of the hypersurface is
given by
\begin{equation}
\rho(a, a') = \sum_b \psi (b, a) \psi^* (b, a') \>.
\end{equation}

Since we argued that no causal signal from the hypersurface $T=0,
X<0$ can enter quadrant I, the complete set of states on the
hypersurface $T=0, X>0$ is in fact the complete set of states needed
to describe physics in Rindler space for all time.  Thus, the above
construction is precisely the density matrix used by a Rindler
observer in quadrant I.

Now consider performing this decomposition for the Minkowski ground
state $| 0 \rangle$.  Given some arbitrary set of fields, which we
will denote by $\phi$, we can represent the ground state wave
functional $\Psi_0 (\phi)$ by using the Feynman-Kac formula,
\begin{equation}
\Psi_0 (\phi) = \int_{\cal F} [d A] e^{-I[A]} \>,
\end{equation}
where $I$ is the Euclidean action for the field and the integral is
over the set $\cal F$ of functions defined for $T \ge 0$ and which
match $\phi$ at $T = 0$.  Let $H_R$ denote the Rindler space
Hamiltonian for the field $\phi$, which generates translations in
$\omega$.  Since $\omega$ corresponds to hyperbolic boost angle in
the $(T, X)$ plane of Minkowski space, when we Wick rotate to
Euclidean space, $H_R$ becomes the generator of rotations in the $(T,
X)$ plane.  Thus we can write
\begin{equation}
\Psi_0 (\phi) = {}_L \langle \phi_L | \exp ( - H_R \pi ) | \phi_R
\rangle_R \>.
\end{equation}
Thus the density matrix for the Minkowski vacuum, $\rho_0 (\phi_R,
\phi_R')$, is given by
\begin{eqnarray}
\rho_0 &=& \sum_{\phi_L} {}_R \langle \phi_R | e^{-\pi H_R} | \phi_L
\rangle_L {}_L \langle \phi_L | e^{-\pi H_R} | \phi_R' \rangle_R
\nonumber \\ {}
& & \nonumber \\ {}
&=& {}_R \langle \phi_R | e^{-2\pi H_R} | \phi_R' \rangle_R \>,
\end{eqnarray}
and so the density operator for the Minkowski space vacuum is $\rho_0
= \exp ( -2\pi H_R )$, which is indeed thermal, with inverse
temperature $\beta = 2\pi$.

This phenomenon is known as the Unruh effect: accelerating observers
experience thermal radiation.  But is there any sense in which the
Rindler observer is {\sl
actually} experiencing a bath of thermalized particles?  Would a real
thermometer measure a temperature?  The answer is yes.  Consider the
fact that there are always fluctuations of the vacuum.  These
fluctuations can be described as loops in spacetime.  Some of these
loops will encircle the origin, lying partially inside and partially
outside quadrant I.  But since these loops intersect the surfaces
$\omega = -\infty$ and $\omega = \infty$, as far as the Rindler
observer is concerned, they are particles which are present for all
time.  The Rindler observer sees these fluctuations as a bath of
thermal particles which are ejected from the horizon infinitely far
in the past and which will eventually fall back onto the horizon in
the infinite future.

Note however that this is an interpretation of a particular
phenomenon by a particular observer.  A freely falling inertial
observer would not describe these vacuum fluctuations in the same
way.  In fact, the freely falling observer would not be able to
distinguish these fluctuations from any other vacuum fluctuations,
and would notice nothing out of the ordinary.  It is only the Rindler
observer who can distinguish this thermal radiation, and only the
Rindler observer who must describe the vacuum using a thermal density
matrix.

The proper temperature measured by a Rindler observer at distance
$\rho$ from the horizon can be obtained from the Rindler temperature
$T_R = 1/2\pi$ by using the transformation between Rindler time and
proper time.  The proper temperature is thus given by $T_{proper} =
1/2\pi\rho$.  A Rindler observer will therefore describe the region
close to the horizon as a very hot place, and in order to describe
physics in this region, he will have to understand physics at
extremely high temperatures.  If the Rindler observer uses an
effective theory with some cutoff mass $\Lambda$, then it is natural
to impose a cutoff at a distance of order $1/\Lambda$ from the
horizon, beyond which the observer cannot penetrate.  For
consistency, however,
this boundary surface must be endowed with some set of degrees of
freedom which represent the degrees of freedom integrated out to
obtain the effective theory at scale $\Lambda$, and should behave
like a hot membrane.  This surface effectively augments the horizon,
and is known as a {\sl stretched horizon}.  We will return later to
this very important idea.

Now that we have determined that a Rindler observer experiences a
temperature which decreases as one moves away from the horizon, let
us return to the quantum fields.  For each transverse momentum mode
$\kv \ne 0$, the field $\phi_{\kv}$ is excited to a thermal spectrum.
 Each mode is populated according to the Boltzmann distribution with
temperature $1/2\pi$, so the modes with energies greater than
$1/2\pi$ will be exponentially suppressed.  But for energies less
than $1/2\pi$, there will exist a bath of thermal particles which
create a thermal atmosphere around the black hole.  Only for the mode
with $\kv = 0$ can these particles escape to infinity.

As we saw earlier, Rindler space is a good description of any region
of the horizon small compared to the entire black hole.  Let us then
consider the interpretation of these results for the finite mass
black hole.  High angular momentum particles which are ejected from
the region close to the horizon are deflected by the centrifugal
barrier and become part of the thermal atmosphere of the black hole.
Only the lowest angular momentum modes can escape the centrifugal
barrier.  This slow leakage of particles out of the centrifugal
barrier is known as {\sl Hawking radiation} \cite{hawk}, and leads to
the eventual evaporation of the black hole.  The temperature as
observed by an observer at asymptotic infinity is $T_{\infty} = T_R
d\omega/dt = 1 / 8 \pi GM$, which is known as the Hawking
temperature.  This implies an evaporation time for the black hole of
order $G^2 M^3$.  The fact that only the lowest angular momentum
modes can escape is the reason for the long evaporation time
of the black hole.

\section{Gedanken Experiments Involving Black Holes}

In the previous sections, we have argued that an external observer
can describe the black hole as a hot membrane which can absorb and
emit particles.  Let us now consider a gedanken experiment designed
to test the existence of the stretched horizon \cite{st}.

Suppose that physics below some energy scale $\Lambda$ can be
described by a more-or-less standard grand unified theory.  If the
claim that the stretched horizon behaves like a hot membrane is
correct, then an observer near enough to the horizon ought to be able
to detect baryon number violation.  Now, we can make the black hole
as large as we like--galactic size, for example--so all the tidal
forces at the horizon are exceedingly small, and it is hard to
imagine how we are going to see any baryon number violation.
Nevertheless, let us press on and see what we find.

Imagine constructing what we call a ``GUT bucket''.  The GUT bucket
is sealed, so that no GUT particles can enter or leave the container,
and has the property that it can withstand GUT scale temperatures,
but not Planck scale temperatures.  Although we know of no way to
construct such a GUT bucket, there is no reason to think that such a
container in any way violates the laws of physics, so it should be
perfectly fine to perform gedanken experiments with it.  We can
imagine that the bucket is suspended above the black hole by some
mechanism like a winch, and that we can slowly lower the bucket
toward the surface of the black hole.

The experiment we will perform, then, is to start off with the bucket
far away from the horizon, where temperatures are far less than the
GUT scale, with no baryons in it.  We will then slowly lower the
bucket down toward the horizon until it has reached a distance of
order $1/\Lambda$, and after some time, slowly raise the bucket back
to the region far from the horizon.  We will then open the bucket and
find out if baryon number has been violated.

If we do this many times, on the average we will find that baryon
number has not been violated, but for any given individual trial, we
can expect to see baryon number violation.  The theory behind this is
quite simple.  In order to suspend the GUT bucket a distance
$\Lambda$ from the horizon, we must subject the GUT bucket to an
acceleration of order $1/\Lambda$ to prevent it from falling into the
black hole.  This acceleration disturbs the vacuum within the bucket
on frequency scales of order $\Lambda$, and the interactions of these
vacuum fluctuations with the walls of the bucket, {\sl etc.} can
produce baryon number violating effects.

Suppose we now change the experiment a little: suppose we allow the
bucket to fall through the horizon freely.  As we stated earlier,
there are no large gravitational or tidal forces at the horizon of a
big black hole, so there is no local signal that the bucket has
fallen through the horizon, and an observer in the bucket will not be
able to detect any baryon number violation.  On the other hand, once
the bucket has fallen through the horizon, it cannot report to the
outside world that there was no baryon number violation, so there is
no contradiction here.  An outside observer cannot derive a
contradiction because he cannot get the information that there was
no baryon number violation, and the observer inside the bucket
detects no baryon number violation until he meets his demise at the
singularity.

So, can we change the experiment to try and obtain a contradiction?
Suppose we prepare another GUT bucket, which we will allow to fall
freely through the horizon, but this time we program the bucket to
send out a message when it has reached a distance $1/\Lambda$ from
the horizon to tell us whether or not there is any baryon number
violation.  If, however, the experimental apparatus and signalling
device all operate by sending out electromagnetic radiation (i.e.
they work as fast
as they possibly can), then the entire process of measurement and
signalling must occur within a time interval $1/\Lambda$.  This
process introduces frequencies of order $\Lambda$ into the
measurement and signalling processes, so the very act of measuring
and reporting on the baryon number within the bucket in a time
interval of order $1/\Lambda$ can, in fact, change the baryon number
within the bucket.  This is a classic example of quantum mechanical
complementarity.

There is a second and perhaps more convincing way to understand the
results of the above experiment.  Let us for the moment igore
questions of confinement, and suppose that we are able to put a
single quark into the GUT bucket.  Let us suppose further that the
GUT theory has dimensionless coupling $\alpha$ at the scale
$\Lambda$, which governs the strength of an interaction between
quarks, antiquarks, and X-bosons.  This is a baryon number violating
interaction.  In what sense, then, is baryon number conserved?  If we
were to calculate the fraction of a given interval
of time that a quark spends as an anti-quark, that is, the fraction
of the time spent ``in the wrong state'', we would find that this
fraction is non-negligible, and is in fact proportional to $\alpha$.
The point is, however, that these virtual transitions occur very
fast--on time scales of order $1/\Lambda$.  If we define a time
averaged baryon number, where we average over intervals large
compared to $1/\Lambda$, then we would find that it is this averaged
or effective baryon number
which is very hard to change.  It is in this sense that the baryon
number violating
interactions are unimportant for low energy physics, because we need
high energy interactions to notice them.

Now suppose that the quark in our GUT bucket makes such a transition
as it is passing through the horizon.  This will occur in roughly a
fraction $\alpha$ of the trials.  Thus, during the entire time at
which the measurement is occuring, the quark is in the wrong state,
and the measurement apparatus will report this result.  We see,
therefore, that it is the uncertainty principle of quantum mechanics,
and nothing to do with gravitational forces, that causes the
experiment to report baryon violation.

A third way of understanding this is to draw the Feynman diagram of a
quark undergoing this transition.  If the loop representing the
virtual states does not lie entirely within quadrant I of Rindler
space, then the virtual states necessarily intersect either $\omega =
-\infty$ or $\omega = \infty$.  Thus, from point of view of a Rindler
observer, the transition is not virtual at all, but represents an
interaction which changes a quark into an anti-quark and X-boson,
which remain in the Rindler spacetime for all time.

There are other gedanken experiments one can devise to test the
existence of the stretched horizon from the external observer's point
of view \cite{st}.  In general, one finds that the description is
consistent with the known laws of physics, and any apparent paradoxes
can be traced to unwarranted assumptions about the nature of physics
at the Planck scale.

\section{The Stretched Horizon and Black Hole Complementarity}

Throughout the last few sections, we have developed some new insight
into the way a black hole can be described by an external observer.
This description is in terms of a stretched horizon, which we
understand loosely as a very hot membrane just above the black hole
surface.  From the point of view of an external observer, the
stretched horizon can interact with infalling matter, and can absorb
and thermalize information possessed by that matter.  Let us now
summarize what is known about the stretched horizon.  Much of this
was developed in the book by Thorne, Price, and
MacDonald \cite{tpm}.

The stretched horizon can be summarized as a membrane which lies just
above the event horizon.  In the formulation of \cite{tpm}, the exact
distance of this membrane above the horizon is somewhat arbitrary,
and is generally fixed for convenience.  For example, if the outside
observer performs measurements which are sensitive to energies up to
the weak scale, then the stretched horizon may profitably be thought
of as lying at a distance of about the weak scale from the event
horizon.  The interactions of the stretched horizon with the outside
world can be thought of as arising from the boundary conditions that
must be implemented in the cutoff theory.  More recently, studies of
black hole horizons in the context of string theory have shown that
the stretched horizon is most naturally thought of as lying at the
string scale above the event horizon, where the local temperature
becomes of order the Hagedorn temperature, as seen below.

The temperature of the horizon is given by $T_{SH} = 1/2 \pi
\rho_{SH}$, where $\rho_{SH}$ is the proper distance from the event
horizon to the stretched horizon.  This is a universal temperature
for all non-extremal black holes for a given $\rho_{SH}$.  Remember
that we found that the centrifugal barrier causes almost all of the
particles emitted from the black hole to be deflected back in, so
these particles form a thermal bath above the black hole horizon.
The fact that very few of the particles get out means that the black
hole evaporation process is very slow,
so the idea of thermal equilibrium and temperature is well defined.

The horizon has other interesting physical properties as seen by an
external observer.  It has an electrical resistivity.  If you place
two electical leads on the stretched horizon and measure the
resistance, you find a resistivity of 377 ohms/square.

The horizon has an entropy per unit area, given by the
Bekenstein-Hawking formula \cite{bek,hawktwo}
\begin{equation}\label{eqn:horent}
\frac{S}{A} = \frac{1}{4G} \>.
\end{equation}
This formula is easily obtained from the thermodynamic relation $dE =
TdS$, where the energy $E$ of the black hole is simply its mass $M$,
the temperature is given by $T = 1/8 \pi GM$, and the area of the
horizon is $A = 16 \pi G^2 M^2$.  (There is a constant of integration
which comes from this formula, which gives a subleading term which
will not concern us much here.  We are not going to entertain the
possibility that this constant is infinite.)

The horizon also has an energy per unit area,
\begin{equation}\label{eqn:horenergy}
\frac{E}{A} = \frac{1}{8 \pi G \rho_{SH} } \>,
\end{equation}
which is obtained as follows.  The Poisson bracket between
Schwarzschild time and energy is given by $\{ t, M \} = 1$.  Now
consider an observer at rest very close to the horizon.  The proper
time of an observer at proper distance $\rho$ from the horizon is
given to leading order in $\rho$ by
\begin{equation}
\tau = \frac{t}{4GM\rho} \>.
\end{equation}
Let the energy of the black hole as measured by this observer be $E$.
 Then we have the Poisson bracket relation $\{ \omega, E \} = 1$.
Writing $E(M)$, we obtain the relation
\begin{equation}
\{ t, E(M) \} = 4GM\rho \>.
\end{equation}
Integrating up this equation gives
\begin{equation}
E(M) = 2GM^2 \rho \>,
\end{equation}
and thus $E/A = 1/8 \pi G \rho$.  Setting
$\rho = \rho_{SH}$ gives the result Eq.~(\ref{eqn:horenergy}).

The horizon also has both bulk and shear viscosity.  If an object
falls into black hole, it will deform the surface of the black hole,
and the deformation will propagate around the surface as would a
disturbance in a viscous liquid.  So, it is starting to sound like
the properties of the horizons of all non-extremal black holes are
universal, and indeed this is so.  There is something strange about
all this, though, in view of the fact that a freely falling observer
will not
see any of this.  The conclusion is that there is no invariance to
the existence of the stretched horizon--its existence depends upon
one's state of motion.

We would now like to introduce a principle, called the {\sl Principle
of Black Hole
Complementarity.}  This is a speculative principle which was
introduced to solve the black hole information problem \cite{stu}.
The principle can be stated as follows.

\begin{itemize}

\item{From the point of view of an external observer, the stretched
horizon exists and is a collection of quantum mechanical, microscopic
degrees of freedom which can absorb, store, thermalize, and emit any
quantum mechanical information which falls into the black hole.}

\item{A freely falling observer will not detect the stretched
horizon, nor will he experience any other local signal when he
crosses the horizon.}

\end{itemize}

We saw above in Chapter~4 that there are no obvious internal
inconsistencies in the principle of black hole complementarity,
although there are some serious arguments why it is wrong.  but
before we examine these arguments, let us digress for a moment and
examine the connection between information, entropy, and how normal
thermodynamic systems are supposed to behave.

\section{Information and Entropy}

We have not really defined information yet, although we have alluded
to the paradox of information loss in black hole evaporation.  Let us
first consider the concept of entropy of entanglement between two
quantum systems.  Suppose we have two quantum systems, $A$ and $B$,
and suppose the combined system is in the state $| \psi \rangle \in
{\cal H} = {\cal H}_A \otimes {\cal H}_B$.  As we saw in section 3,
if we trace over the degrees of freedom in ${\cal H}_B$, we obtain a
density operator $\rho$ whose matrix elements are given by
\begin{equation}
\rho_A (a, a') = \sum_b \psi (a, b) \psi^* (a', b) \>.
\end{equation}
Given such a density matrix, one can obtain the entropy of
entanglement between $A$ and $B$, which is defined to be
\begin{equation}\label{eqn:sent}
S_A = -\tr \left ( \rho_A \ln ( \rho_A ) \right ) \>.
\end{equation}
It is easy to see that $S_A = S_B$.  The entropy of entanglement is
essentially the logarithm of the number of independent states which
have a nontrivial probability of being occupied.  This entropy arises
because the systems have non-trivial correlations with each other.

There is another kind of entropy, which we will call entropy of
ignorance.  This entropy is not there because the system is
necessarily entangled, but simply because you have not measured all
the possible variables needed to completely specify the state of the
system.  For example, thermodynamic entropy is of this type.  For a
thermal system, instead of using a density operator obtained from
tracing out microscopic degrees of freedom, one simply postulates a
density operator of the form $\rho = \exp ( - \beta H ) / Z(\beta)$,
where $Z(\beta)$ is the partition function.  In general,
$S_{entanglement} \le S_{thermal}$, because $S_{thermal}$ represents
not only the entanglement of the system with its environment, but
also entropy our ignorance of it.  The information $I$ contained in a
system can be defined as the difference between the maximum entropy
of ignorance, usually taken to be the thermodynamic entropy, and the
entanglement entropy, so $I = S_{max} - S_E$.

To understand how entropy and information behave in ordinary systems,
let us now consider a gedanken experiment due to Sidney Coleman.
Consider a lump of black coal, which we will treat as an ideal black
body, at zero temperature.  Let us illuminate the coal with a
sequence of pulses from a laser beam.  This heats up the coal, which
begins to glow and radiate away the energy absorbed from the laser in
the form of thermal radiation.  This continues until the coal has
cooled back down to its ground state.  Since we know that the
$S-$matrix for this process exists and is unitary, it must be true
that the information coded in the sequence of pulses must still be
present.  Since the coal has cooled back down to zero temperature, it
contains no information, so the information must be contained in the
radiation field.  So we see that thermal radiation can, in fact, code
information, although we have not determined the mechanism by which
it does this.

\begin{figure}
\vbox{{\centerline{\epsfsize=88cm \epsfbox{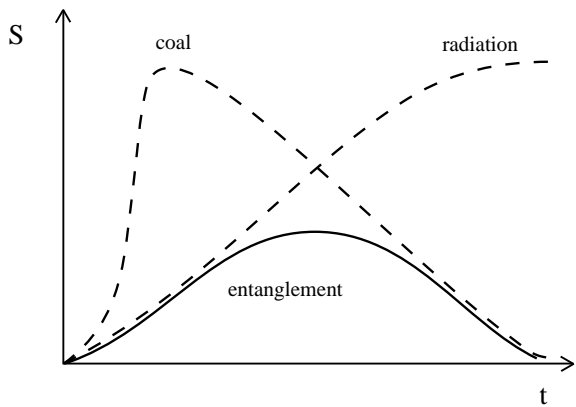}}}}
\caption{Plot of thermal and entanglement entropies of two
subsystems.}
\label{fig:entropyplot}
\end{figure}

This question was analyzed in a brilliant paper by Don Page
\cite{dp}, whose results imply that the information is in fact coded
in long-time, long-distance correlations between the photons.  The
analysis went something like this.  Consider the combined quantum
system composed of the radiation field and the lump of coal.
Initially, the thermal entropy of the coal is very large, while that
of the radiation field is very small.  In Fig.~\ref{fig:entropyplot}
we have plotted the thermal and entanglement entropies of the systems
as functions of time.  The entanglement entropy, common to both
systems, must be less than either of their thermal entropies.  Page
found that the entanglement entropy is almost exactly equal to the
thermal entropy of the radiation field, up to the point where the
thermal entropies of the radiation and the coal are equal.  At the
crossover point, the entanglement entropy comes within about one bit
of the thermal entropy, and then begins decreasing.  It closely
follows the thermal entropy of the coal down to zero.  At this point,
the systems are completely unentangled, but there is a large amount
of thermal entropy (i.e. entropy of ignorance) in the radiation
field.  The information represented by this entropy contains the
information coded in the sequence of laser pulses.

The question Page asked was, how much of the combined system must we
measure before we can get any information out?  The answer is that we
must be able to sample at least half of the combined system before we
can obtain even one bit of information.  This shows that the
information is not contained in short distance correlations, which
could be distinguished by examining small portions of the system, and
so must be contained in long-distance correlations.  From this, we
see that the information in the laser beam is first stored by the
coal, and then transferred to the radiation field in the form of
long-distance (or long-time) correlations between the photons.

Of course, we are interested in black holes, not coal.  The point is
that the principle of black hole complementarity implies that black
holes behave exactly the same way.  The information which is
contained in the infalling matter is thermalized by the stretched
horizon, and re-emitted in the Hawking radiation.  But to decode the
initial information, one must look at very long time correlations
between the Hawking photons.  To retrieve even one bit of the initial
information, even in principle, one must wait for a time of order
half the lifetime of the black hole.  In reality, decoding the
information from the radiation field is an extraordinarily
complicated process (as in the case of the coal).

\section{Formal Arguments Against Black Hole Complementarity}

The first argument we shall consider concerns the location of
information in the plane of the horizon.  Suppose we have a large
black hole, and suppose we allow a neutron to fall through the
horizon at transverse location $(\theta_0 , \varphi_0 )$.  Standard
calculations in quantum field theory show that, after the neutron has
fallen through the stretched horizon, the information needed to
distinguish the particle as a neutron ({\sl i.e.} $SU(3) \times SU(2)
\times U(1)$ quantum numbers,
baryon number, lepton number, {\sl etc.}) can be obtained by
measuring a small region around $(\theta_0 , \varphi_0 )$.  This
certainly contradicts the idea that the stretched horizon contains
microscopic degrees of freedom which thermalize and scramble all the
infalling information.

If black hole complementarity is to be correct, then it must be the
case that some new physics enters which thermalizes the information
contained by the neutron, and causes it to spread out over the
horizon.  This is a necessary condition for a theory which implements
black hole complementarity.  We will return to this condition in the
context of string theory shortly.

A second argument concerns the longitudinal behavior of matter as it
falls toward the horizon.  In Chapter 5 we argued that the horizon
has an entropy per unit area given by Eq.~(\ref{eqn:horent}), and in
Chapter 6 we argued that the entropy should be interpreted as the
logarithm of the number of accessible microstates of the system.  Now
consider sending a particle toward a black hole.  In the frame of an
external observer, the momentum of the particle increases
exponentially with
time.  According to the usual Lorentz contraction formula, the fields
associated with the particle will contract in the longitudinal
direction (direction of motion), and this contraction will proceed
indefinitely, since the relative boost increases without bound.  Thus
the longitudinal region occupied or influenced by the particle can be
made arbitrarily small.

In the frame of the external observer, no particle ever crosses the
horizon.  Since we have argued that the longitudinal extent of
particles can be made arbitrarily small, we see that, in effect, we
can stack an arbitrarily large number of particles in layers of ever
decreasing thickness onto the horizon.  This means that the number of
accessible states per unit area, and thus the entropy per unit area,
can be made arbitrarily large, in contrast to the finiteness of the
entropy per unit area given by Eq.~(\ref{eqn:horent}).  Thus, if the
entropy per unit area is to have any type of state counting
interpretation, there must be a mechanism which halts the Lorentz
contraction as some point, and prohibits us from stacking an
arbitrarily large amount of information into the horizon.  This
provides another necessary condition for a theory to implement black
hole complementarity.

We will now give an argument for information loss based on effective
low energy field theory, called the {\sl nice slice argument.}  The
precise formulation of the argument which we will use is due to Joe
Polchinski, but the argument is implicit in the literature dating
back many years.

Consider the geometry for the formation of a large black hole by
infalling matter, and imagine foliating the geometry by a family of
Cauchy surfaces that have the following properties.  First, the
momenta of all the infalling particles are small in the frame of the
surfaces.  Second, the outgoing Hawking radiation has small momentum
in the frame of the surfaces, up until the point that the black hole
has become so small that the semiclassical approximation breaks down.
 Finally, the surfaces are everywhere smooth, so that there are no
large local geometric quantities.  We will call such a family of
Cauchy surfaces a family of {\sl nice slices.}  The first
construction of a family of nice slices that we are aware of is due
to Wald.

It is now easy to give the argument for information loss.  Since all
momenta relevant to the black hole formation and evaporation
processes are small in the nice slice frame, and the adiabatic
theorem guarantees that all the high energy modes will be in their
ground states, we can use low energy effective field theory to
describe the process.  Consider some information encoded in local
correlations of the effective fields which propagates into the black
hole.  The equivalence
principle states that nothing out of the ordinary happens to the
infalling wave packet as it crosses the horizon, so a local observer
falling with it must necessarily be able to retrieve the information
after he has entered the black hole.  This shows that the information
is certainly inside the black hole.

We now invoke the so-called ``no quantum Xerox principle''.  Suppose
that the Hilbert space of states ${\cal H}$ factorizes into ${\cal
H}_{in} \otimes {\cal H}_{out}$, and suppose that we try to build a
quantum Xerox machine that replicates the information contained in a
state $| \psi \rangle_{in} \in {\cal H}_{in}$ into a state in ${\cal
H}_{out}$.  The quantum Xerox machine must be a linear operator on
the Hilbert space, which we will denote $X$.  Then our rule is
\begin{equation}
X \left [ | \psi \rangle_{in} \otimes | \gamma \rangle_{out} \right ]
= | \psi \rangle_{in} \otimes |
\psi \rangle_{out}
\end{equation}
for any state $| \gamma \rangle_{out} \in {\cal H}_{out}$.  Now we
can always write $| \psi \rangle_{in} = | \alpha \rangle_{in} + |
\beta \rangle_{in}$ for some $| \alpha \rangle_{in}$ and $| \beta
\rangle_{in} \in {\cal H}_{in}$.  But then
\begin{eqnarray}
&& X \bigl [ | \psi \rangle_{in} \otimes | \gamma \rangle_{out} \bigr
] \nonumber \\
&=& X \left [ \left ( | \alpha \rangle_{in} + | \beta \rangle_{in}
\right ) \otimes | \gamma \rangle_{out} \right ] \nonumber \\
&=& X \left [ | \alpha \rangle_{in} \otimes | \gamma \rangle_{out}
\right ] + X \left [ | \beta \rangle_{in} \otimes | \gamma
\rangle_{out} \right ] \nonumber \\
&=& | \alpha \rangle_{in} \otimes | \alpha \rangle_{out} + | \beta
\rangle_{in} \otimes | \beta \rangle_{out} \nonumber \\
&\ne& | \psi \rangle_{in} \otimes | \psi \rangle_{out} |>.
\end{eqnarray}
Thus we see that the operator $X$ is not in fact a linear operator,
and so is not admissible.  This shows that there is no such thing as
a quantum Xerox machine.

Now let us show why information must be lost.  Since we are able to
describe the black hole evaporation process using effective local
fields which commute at spacelike separation, the Hilbert space of
the effective field theory factorizes into the product ${\cal H}_{in}
\otimes {\cal H}_{out}$.  The no quantum Xerox machine principle then
applies, stating that if the information contained by infalling
matter is certainly contained within the black hole, then it is
certainly
not contained in the region outside the black hole.  Moreover, since
the fields commute at spacelike separation, there is no way for the
information to get outside the black hole once it has entered.  Thus,
the information cannot escape and is lost to the external observer.

This seems to be an air-tight argument for information loss, but it
includes a crucial assumption, which is that the low energy theory is
in fact a local field theory.  That this is the case is not at all
obvious, especially in string theory, so there is still hope.  In the
next section we begin our examination of string theory, where we will
see that there are indications that black hole complementarity is
realized in string theory.

\section{Light-Front String Theory and Complementarity}

String theory is a theory of strings, not particles, and it behaves
very differently from ordinary quantum field theories in many
respects.  In this section we will analyze the physics of stringy
matter falling toward a horizon as described by an external observer,
and will see that there is circumstantial evidence that black hole
complementarity is realized in string theory.  Specifically, we will
show that the necessary conditions of transverse spreading and the
cessation of longitudinal Lorentz contraction both occur in string
theory.

Although we presently do not have the technology to quantize string
theory in a black hole background, we saw earlier that in the
vicinity of the horizon, the black hole geometry is well approximated
by Rindler space, so we shall use that approximation.  Then we simply
have strings in flat space, which we understand how to deal with.

We will use the light-front gauge formalism of string theory.  There
are a number of reasons why the light-front formalism is convenient
for this problem.  Imagine a single string falling through a Rindler
horizon.  Since the surfaces of constant Rindler time $\omega$ are
obtained by boosting the surface $T = 0$, it is clear that, given any
inertial frame, the surfaces of constant $\omega$ become almost
light-like with respect to this frame for $\omega$ large enough.
Thus, evolution in Rindler time for large $\omega$ is well
approximated by evolution in light-front time.

An important fact to notice is that the relation between proper time
$\tau$ in the frame of the string and Rindler time $\omega$ is given
for $\omega$ large by
\begin{equation}\label{eqn:dtaudomega}
\frac{d\tau}{d\omega} = \rho e^{-\omega} \>.
\end{equation}
Therefore, as $\omega$ gets large, less and less proper time in the
frame of the string elapses per unit of Rindler time, so a Rindler
observer ``sees'' the string with an ever increasing time resolution.
 More precisely, if a Rindler observer at fixed proper distance from
the horizon samples outgoing photons at a uniform rate of one sample
per unit Rindler time, then the world lines of the sampled photons
form a sequence of light like surfaces which accumulate at the
horizon at the rate given by Eq.~(\ref{eqn:dtaudomega}).

There is another way to think of this.  We have already seen that the
momentum of an infalling string, as reckoned by an external observer,
increases like $e^{\omega}$ as $\omega$ gets large.  The increased
time resolution can simply be thought of as the Lorentz contraction
of time between the two frames.

Let us now review some of the fundamentals of light-front gauge
string theory.  We will work in units where $\alpha' = 1/2$.  The
world sheet coordinates are taken to be $(\tau , \sigma )$, where
$\tau$ is the timelike direction, and the metric on the world sheet
is just the usual two-dimensional Minkowski metric.  We define light
front coordinates in spacetime by
\begin{equation}
X^{\pm} = \frac{1}{\sqrt{2}} \left ( X^0 \mp X^{D-1} \right ) \>,
\end{equation}
and choose the gauge $X^+ = \tau$.\footnote{Note that the usual
convention is $X^{\pm} = \frac{1}{\sqrt{2}} ( X^0 \pm X^{D-1} )$, but
the above choice is more convenient for our purposes.}  After gauge
fixing, the longitudinal coordinate $X^-$ is completely determined in
terms of the transverse coordinates $\Xv$, and the action for the
transverse coordinates is simply
\begin{equation}
S = \frac{1}{2 \pi} \int d\tau d\sigma \left [ \left ( \pa_{\tau} \Xv
\right )^2 - \left ( \pa_{\sigma} \Xv \right )^2 \right ] \>.
\end{equation}
In addition, we have the condition that $p^+$, the longitudinal
momentum, is conserved, and is uniformly distributed along the
string.  For convenience, we will set $p^+ = 1$.  The transverse
string coordinates can be expanded as
\begin{equation}\label{eqn:tcexpand}
\Xv (\sigma) = \xv + \frac{i}{2} \sum_{n \ne 0} \frac{1}{n} \left [
\av_n e^{2in\sigma} + \tilde{ \av_n } e^{-2in\sigma} \right ]
\end{equation}
where $\xv$ is the center of mass position of the string at $\tau =
0$ and the $\alpha^i_m$ obey the commutation relations
\begin{equation}\label{eqn:comrel}
[ \alpha^i_m, \alpha^j_n ] = m \delta^{ij} \delta_{m + n} \>.
\end{equation}

Let us now use light-front string theory to answer the question, how
big is a string?  Specifically, we will calculate the mean square
transverse size of a closed bosonic string in its ground state.  This
is given by the matrix element
\begin{equation}\label{eqn:msmel}
\langle 0 | \left ( \Xv (0) - \xv \right )^2 | 0 \rangle \>.
\end{equation}
Explicit calculation of Eq.~(\ref{eqn:msmel}) using the mode
expansion Eq.~(\ref{eqn:tcexpand}) yields
\begin{equation}\label{eqn:msdiv}
\langle 0 | \left ( \Xv (0) - \xv \right )^2 | 0 \rangle =
\frac{D-2}{2} \sum_{n=1}^{\infty} \frac{1}{n}
\end{equation}
which diverges logarithmically.  Thus our calculation leads to the
idea that the string has an infinite transverse extent.  But what
does this mean?  Is this another infinity which should be
renormalized somehow?  The answer is no - let us see why.  Consider
an actual experiment designed to measure the transverse size of a
string.  Such an experiment will take some amount of time
$\varepsilon$ to perform, and thus will be insensitive to frequencies
greater than $N \sim 1/\varepsilon$.  Thus the mode sum in
Eq.~(\ref{eqn:msdiv}) should be cut off at $N$, giving a finite but
$\varepsilon-$dependent answer,
\begin{equation}\label{eqn:msfin}
\langle 0 | \left ( \Xv (0) - \xv \right )^2 | 0 \rangle =
\frac{D-2}{2} \sum_{n=1}^N \frac{1}{n} \>,
\end{equation}
which behaves like $\log ( 1/\varepsilon)$.  We see that as the
resolution time $\varepsilon$ goes to zero, the measured size of the
string diverges logarithmically.  This behavior is nothing more than
the Regge nature of string scattering amplitudes.  It is also
possible to show that the total measured length of string diverges
like $1/\varepsilon$ \cite{kks}.  From this we see that as
$\varepsilon$ decreases, the string appears to fill space more and
more densely.

A similar calculation can be performed for the longitudinal size of
the string, and one finds that \cite{sbhlc}
\begin{equation}
\langle 0 | \left ( X^- (0) - x^- \right )^2 | 0 \rangle \sim
\frac{1}{\varepsilon} \>.
\end{equation}
This shows that the usual Lorentz contraction along the direction of
motion is eventually halted in string theory, and the string begins
to grow like $1/\varepsilon$.

Now let us return to the case of an external observer watching a
string fall toward the event horizon of a large black hole.  From
Eq.~\ref{eqn:dtaudomega}, we saw that an external observer has a time
resolution that decreases like $e^{-t/4GM}$.  Thus, an external
observer will see the mean squared radius of the string grow like
$t/4GM$, while its total length and longitudinal spread will increase
like $e^{t/4GM}$.

Using this, we see that the time to spread over the entire horizon is
of order $t_{\hbox{\scriptsize spread}} = G^3 M^3$, which is much
less than the evaporation time $t_{\hbox{\scriptsize evap}} = G^2
M^3$.  In addition, the longitudinal spread causes the string to fill
up a region near the horizon, where it appears to float.

The previous analysis shows that someone observing an infalling
string will find that it appears to grow and cover the horizon,
forming a sort of stringy goo just above the horizon \cite{mpt}.  One
may rightly ask if there is any sense in which the information
contained in the string state also is spread out, or thermalized.  In
ordinary field theory, degrees of freedom are independent of each
other unless they have a non-vanishing commutator.  It is then
natural to ask, what is the analogous statement in string theory?

In work by Lowe and the authors \cite{lsu}, this question is
addressed using light front bosonic string theory.  The commutator of
two component fields at equal light front time is shown to be
non-vanishing to first order in the string coupling even when the
arguments of the fields are spacelike (transversely) separated.  This
shows that the information in the string state also spreads.

In more recent work by Lowe, Polchinski, Thorlacius, and the authors
\cite{lpstu}, the commutator of appropriately dressed low energy
component fields is calculated and shown to be nonzero when the
fields are at different light front times.  More precisely, suppose
observer 1 at position $x_1$ stays in front of the horizon at some
finite distance $\rho$, while observer 2 freely falls through the
horizon.  Let $x_2$ lie on the world line of observer 2 behind the
horizon.  Let $\phi(x_1)$ and $\phi(x_2)$ be component fields which
are low energy in the frame of the respective observers.  Then the
matrix element
\begin{equation}
\langle 0 | [ \phi (x_1), \phi (x_2) ] [ \phi (x_2), \phi (x_1) ] | 0
\rangle
\end{equation}
is shown to be nonzero to first order in the string coupling, and in
fact grows like $e^{t/4GM}$.  This growth can ultimately be traced to
the Regge behavior of string scattering amplitudes, and the existence
of the graviton in string theory.

The calculation of the size of a string in its ground state was
performed using free string theory.  We saw that because the length
of string grows faster than the area it occupies, the string becomes
dense as $\varepsilon$ decreases.  When the transverse density
becomes of order $1/g^2$, interactions will become important, and our
perturbative calculation can no longer be trusted.  Presumably, at
this stage non-perturbative string physics takes over and the density
does not grow past the Planck density.  We will have more to say
about this later.

\section{Black Hole Entropy in String Theory}

It was stated earlier that black holes have an entropy, given by the
Bekenstein-Hawking formula,
\begin{equation}\label{eqn:bhent}
S = \frac{A}{4G \hbar} \>,
\end{equation}
where $A$ is the area of the horizon and we have made the factor of
$\hbar$ explicit.  It should be pointed out here that the fact that
black holes have a non-vanishing entropy is not due to quantum
mechanics.  The fact that the entropy is {\sl finite} is due to
quantum mechanics, as can be seen in Eq.~(\ref{eqn:bhent}).  As
$\hbar \rightarrow 0$, the entropy diverges.

The fact that black holes have entropy has been an extremely
confusing one, for at least two reasons.  We ordinarily think of
entropy as having to do with counting the number of degrees of
freedom of a system, and since the entropy is proportional to the
area of the horizon, it certainly seems as if we are counting the
states of something near the horizon.  Precisely what states are
being enumerated, however, remains unclear.  What makes this even
more confusing is that we are attributing an entropy to something
which is essentially a classical solution of the gravitational field
equations, something which seems very much like a soliton, and we
certainly do not ordinarily assign an entropy to a soliton.

On the other hand, it is possible to calculate the entropy of a field
propagating on a fixed black hole background.  We saw in Chapter 3
that the correct way to think of a quantum field propagating outside
a black hole is as a thermal system, so we could follow 't Hooft and
propose that the entropy of the black hole is nothing but the entropy
of the thermal atmosphere of particles around the black hole.  The
trouble is, explicit calculation of the entropy of a massless, real
free scalar field $\phi$ in the thermal or Unruh state outside a
black hole shows that the entropy is
quadratically divergent, and is given by
\begin{equation}\label{eqn:sfent}
S_{\phi} = \frac{cA}{4 \varepsilon^2}
\end{equation}
where $\varepsilon$ is a proper distance cutoff of the theory and $c$
is a numerical factor which depends on the precise form of the
cutoff.  This is in sharp distinction to Eq.~(\ref{eqn:bhent}), which
is manifestly finite.

This raises the question, should we include this divergent entropy as
part of the entropy of the black hole?  If not, why not?  The point
is that the divergent entropy of the field $\phi$ is due to the
enormous number of states that are available to the field near the
horizon.  This can be seen by explicit computation, but it is
connected with the ideas introduced earlier of what happens to matter
as it approaches the horizon. If particles are allowed to Lorentz
contract to arbitarily small longitudinal extent, then an arbitrarily
large number of them can be packed close to the horizon, giving a
divergent entropy.  These states are of extremely short wavelength,
but do not carry a lot of energy because of the redshift phenomenon.
The above paradoxes have led some physicists to abandon the
connection between entropy and state counting, but we shall assume
that it continues to hold for black holes, and see where that leads
us.

Let us now consider how to calculate the entropy of a Rindler
horizon.  We have seen that one of the advantages of Rindler horizons
over real black hole horizons is that we can use flat space, where we
know basically how to perform calculations.  The Rindler horizon is
infinite in area, but we expect the entropy per unit area to be well
defined.

One way of doing statistical mechanics for a canonical ensemble is to
perform a Wick rotation to imaginary time, and then make the time
variable periodic with period $\beta = 1/T$, where $T$ is the
temperature of the canonical ensemble.  Thus we will be dealing with
Euclidean Rindler space, which has the line element
\begin{equation}\label{eqn:erm}
ds^2 = \rho^2 d\theta^2 + d\rho^2 + d\xv^2 \>,
\end{equation}
where $\theta \in (0, \beta )$ is the Euclidean time variable and
$\xv$ denotes the transverse coordinates.  The horizon has now been
compressed to the surface $\rho = 0$.  If $\beta = 2\pi$, then
Eq.~(\ref{eqn:erm}) is nothing more than the line element for flat
Euclidean space in cylindrical polar coordinates.  For $\beta \ne 2
\pi$, the space has the geometry of the cross product of a cone with
$D-2$ dimensional flat space.  This geometry has a curvature
singularity at the horizon.  Since such a singularity was not present
in the original spacetime, we must choose $\beta = 2 \pi$, rederiving
the Rindler temperature.

The Rindler Hamiltonian now generates rotations of the space, and the
Unruh density operator is simply equal to
\begin{equation}
\rho_{Unruh} = \frac{ \exp ( - \beta H_R ) }{ Z ( \beta ) } \>,
\end{equation}
where $Z ( \beta )$ is the partition function,
\begin{equation}
Z ( \beta ) = \tr \left ( \exp ( - \beta H_R ) \right ) \>.
\end{equation}
We now want to calculate the entropy for this density operator with
$\beta = 2 \pi $.

One way of calculating the entropy is to recall the thermodynamic
relations
\begin{eqnarray}
Z ( \beta ) &=& \exp \left ( - \beta F ( \beta ) \right ) \nonumber
\\
\hbox{and} \nonumber \\
S ( \beta ) &=& \beta^2 \frac{ \pa F }{ \pa \beta } \>.
\end{eqnarray}
In order to use these identities, however, we have to be able to vary
$\beta$ away from $2 \pi$, which means we must do physics, either
field theory or string theory, on a cone.

Let us imagine representing the partition function $Z ( \beta )$ as a
functional integral over all geometries with certain boundary
conditions.  The leading order contribution to the functional
integral is simply $\exp ( - I_C ( \beta ) )$, where $I_C$ is the
Euclidean action evaluated for the solution of the classical
equations of motion.  (We are ignoring a possible cosmological
constant term, which cannot contribute to the entropy anyway.)  Thus
we can write, to leading order, $\beta F = I_C$.  The classical
action is (ignoring surface terms, which do not play a role here)
\begin{eqnarray}
I_C &=& \frac{1}{16 \pi G_0} \int d^4 x \sqrt{g} R \nonumber \\
&=& \frac{(2 \pi - \beta ) A}{8 \pi G_0 } \>,
\end{eqnarray}
where $A$ is the area of the horizon and $G_0$ is the bare
gravitational coupling constant.  This then gives an entropy
\begin{equation}\label{eqn:sbare}
S_0 = \frac{A}{4G_0} \>.
\end{equation}
It cannot be emphasized enough that we {\sl do not} have a
statistical interpretation for the entropy here.  There is no mention
of what states are being counted.  This is simply a formal
procedure--we get the correct answer, but we don't really know what
it means \cite{spec}.

If we add a free scalar field $\phi$ to the system, the entropy of
the field can be calculated in the same way.  This was first done by
't Hooft \cite{thooft}, and later by other authors.  The action can
be represented as a sum of first-quantized particle paths, so we can
identify each contribution with a set of paths.  The set of paths
which do not encircle or touch the horizon do not contribute to the
entropy.  To see why, consider a path in this set.  The action of the
(local) path cannot depend on the (global) value of $\beta$ unless
the path is somehow entangled with the surface $\rho = 0$.
Integrating the center of mass of a specific configuration around the
spacetime gives a volume factor proportional to $\beta$, which is
then divided out to obtain the free energy, and the result is thus
independent of $\beta$, and so contributes zero to the entropy.  On
the other hand, paths which do encircle or touch the origin can
contribute to the entropy, and these are the states which are being
counted.  The calculation is easily done, and Eq.~(\ref{eqn:sfent})
is recovered.

If we add this new contribution to the entropy, we would apparently
violate the universality of the Bekenstein-Hawking entropy.  But we
can write the sum in a suggestive form \cite{su},
\begin{equation}
S_0 + S_{\phi} = \frac{A}{4} \left ( \frac{1}{G_0} +
\frac{c}{\varepsilon^2} \right ) \>,
\end{equation}
and remember that the bare coupling constant is renormalized.  We can
argue that what is going on is in fact the renormalization of the
coupling constant by calculating the renormalization of $G_0$ for an
arbitrary geometry and showing that the coefficients are the same,
and explicit calculation in perturbation theory shows that this is in
fact the case \cite{verify,kss}.  So what we have is essentially a
low energy theorem, stating that it is the fully renormalized
gravitational coupling
which enters the Bekenstein-Hawking formula.  In fact, since we do
not yet actually have a quantum theory of gravity, the above can be
viewed as a sort of consistency condition on the theory.  The main
result we should take away from this is that the question of
finiteness of the black hole entropy is intimately entanglement with
the ultraviolet behavior of the theory of quantum gravity, and is
{\sl not} something that can be understood using only low energy
effective field theory.

Conspicuous by their omission from the previous discussion are
graviton loops.  These loops are very complicated, and in fact lead
to disasters.  These disasters are associated with infrared
divergences, which are the manifestation of the Jeans instability.
The nature of these divergences can be understood by making an
analogy to a plasma.  The Boulware vacuum is the vacuum defined by
eliminating all of the thermal particles from the Unruh vacuum--it is
essentially the Rindler Fock space vacuum.  This vacuum must have a
large and negative Minkowski energy, because the Minkowski vacuum $|
0 \rangle$, which corresponds to the Rindler space Unruh vacuum, has
no Minkowski energy.  We must therefore attribute to truly empty
Rindler space a large and negative energy density.  Starting with the
Boulware vacuum, we can fill it up with thermal particles until
reaches the Unruh state, and the energy density vanishes.  This is
somewhat analogous to having a system with a uniform positive charge
density, and filling up the system with electrons until the net
charge is zero.  People who study the statistical mechanics of
plasmas know that you have to be careful in dealing with the Coulomb
force, and one must regulate the infrared tails in order to define
the statistical ensemble.  For example, varying the charge density
away from its background value will create infrared divergences
unless the Coulomb force is cut off.

This analogy suggests the following order of operations in
calculating the entropy of the horizon.  We must first
infrared-regulate the theory, by cutting off the long range
gravitational field.  Having done this, we can vary $\beta$ away from
$2 \pi$ to calculate the entropy.  After obtaining the functional
form of the entropy, we evaluate it for $\beta = 2 \pi$.  Finally, we
take away the regulator, and hope that the answer remains well
defined.

The story is slightly more complicated for string theory \cite{su}.
As we saw earlier, for $\beta \ne 2 \pi$, the geometry is the product
of a cone with flat space, and this background geometry does not
satisfy the conditions for conformal invariance of the world sheet
theory, and the world sheet theory is not ultraviolet finite.  To
define the theory off-shell, we must introduce a world sheet
ultraviolet regulator into the theory.  But as we saw before, this
has exactly the effect of cutting off the spacetime theory in the
infrared!  Thus, by introducing the world sheet regulator we
simultaneously define the theory off-shell and eliminate the Jeans
instability, so we can expect that the statistical ensemble will be
well defined.  Of course, this is a conjecture about how to define
string theory off shell, so we should check that our answers are
independent of this prescription.

So our task now is to calculate the genus expansion of the free
energy of strings in a conical background, which will take the form
\begin{equation}
\beta F ( \beta ) = \sum_{n = 0}^{\infty} g^{2(n - 1)} Z^{(n)} (
\beta )
\end{equation}
where $g$ is the string coupling constant and $Z^{(n)}$ is the
partition function for the two dimensional theory on a world sheet of
genus $n$.  Consider first the genus zero term.  If we imagine using
a lattice regulator, so that the world sheet is composed of $N$
points, then the partition function of the world sheet theory reduces
to a product of ordinary coupled Gaussian integrals.  Because of the
exponential fall off of the integrands, there can be no nonlocal
behavior introduced into the integrals, so on general grounds, we can
expand the partition function in local geometric invariants as
\begin{equation}\label{eqn:pfexpand}
Z^{(0)} = - \frac{1}{16 \pi G_0 } \left [ \int d^4 x \sqrt{g} R + Q
\right ] \>,
\end{equation}
where $Q$ contains all the other terms which enter the expansion.
(We are again dropping any cosmological constant term, which cannot
contribute to the entropy.)  In general, $Q$ will depend in a
singular way on the world sheet regulator and on the conformal
degrees of freedom, but it is important to note that the first term
does not depend on either.  These singular terms can be interpreted
as the residue of the graviton, whose long range field has been
truncated by the regulator.

The contribution from the first term of Eq.~(\ref{eqn:pfexpand}) is
precisely the term we found before, and gives an entropy equal to
Eq.~(\ref{eqn:sbare}).  For the remaining terms, it is easy to argue
that they either vanish by integration by parts, or are proportional
to $(2 \pi - \beta )^2$, and so will vanish when we set $\beta = 2
\pi$.  Thus, the only term which can contribute to the entropy is
the first term, which gives Eq.~(\ref{eqn:sbare}), and we can safely
remove the world sheet regulator.  So the genus zero contribution to
the entropy of the horizon is precisely the Bekenstein-Hawking
entropy, with the bare coupling constant $G_0$.  The analysis goes
through essentially unchanged for the higher genus contributions,
although for higher genus we must also regulate the corners of the
moduli space.

Since the scattering amplitudes of string theory are finite, we are
guaranteed that when we sum up the contributions, the horizon entropy
per unit area is finite order by order in perturbation theory, so
string theory avoids the problem of any infinite contributions to the
entropy.

The answer we obtained for the contribution to the horizon entropy
from the genus zero partition is exactly the same as that obtained
from the classical action of the gravitational field earlier.  In the
latter case, however, we saw that there was no way to identify which
states the entropy was counting.  Let us now see if we can discover
what states we are counting in string theory.

\begin{figure}
\vbox{{\centerline{\epsfsize=66cm \epsfbox{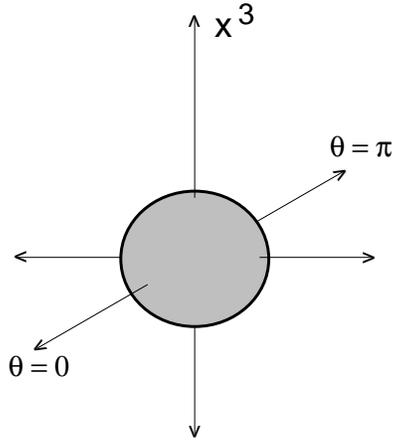}}}}
\caption{A genus zero string graph which contributes to the entropy.}
\label{fig:genuszero}
\end{figure}

As with the field theory case, the only string graphs which can
contribute to the entropy are those which are somehow entangled with
the horizon.  An example for genus zero is shown in
Fig.~\ref{fig:genuszero}.  In order to determine the state which this
graph corresponds to, we simply take a slice of constant Euclidean
time $\theta$.  It is easy to see that the state is described by an
external observer as an open string with both ends frozen to the
horizon.  This state could equally well be described as a closed
string lying partially behind the horizon.  So the entropy
corresponds to the states of fluctuations of open strings pinned to
the horizon surface.

Taking the usual field theory limit $\alpha' \rightarrow 0$, these
configurations degenerate to points on the horizon, and their
interpretation as states is lost.  In this case, we simply have to
say that there is an entropy whose origin cannot be understood from
the low energy theory.

\begin{figure}
\vbox{{\centerline{\epsfsize=66cm \epsfbox{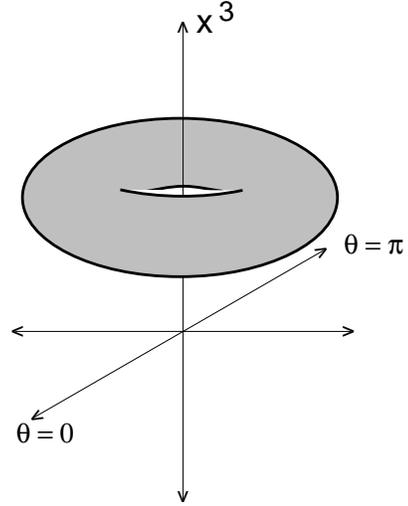}}}}
\caption{A genus one string graph showing a string which remains
permanently outside the horizon.}
\label{fig:genusone-one}
\end{figure}

\begin{figure}
\vbox{{\centerline{\epsfsize=66cm \epsfbox{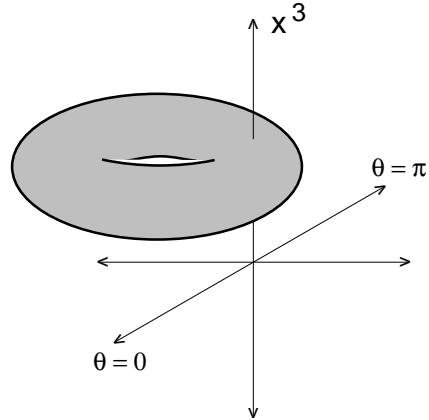}}}}
\caption{A genus one string graph showing an interaction between a
stranded string and an external string.}
\label{fig:genusone-two}
\end{figure}

Fig.~\ref{fig:genusone-one} and Fig.~\ref{fig:genusone-two} show
genus one diagrams which contribute to the entropy.
Fig.~\ref{fig:genusone-one} shows a diagram
whose state-counting interpretation is simply a closed string which
remains outside the black hole.  The $\alpha' \rightarrow 0$ limit of
this diagram simply corresponds to a particle which remains forever
outside the black hole.  Fig.~\ref{fig:genusone-two}, however, cannot
be identified as a single state, but only as an interaction between a
string frozen to the horizon and a string outside the black hole.
This term should not be thought of as the entropy of anything, but as
a correction term.  In the $\alpha' \rightarrow 0$ limit, these
diagrams must be associated with contact terms with the horizon.
These terms have been discovered in ordinary quantum field theory
\cite{kss}.

Thus we see that the stranded strings can interact with each other
and with strings
outside the black hole, so what we have is a complicated many body
problem near the horizon with a large number of degrees of freedom.
This system, as with any other complicated system, is capable of
storing, thermalizing, and re-emitting information which it comes
into contact with.

\section{The Holographic World}

In the previous chapters, we have seen that strings exhibit a variety
of new and interesting phenomena which are relevant to the physics of
black holes.  These include the observed transverse spreading, the
reversal of longitudinal Lorentz contraction, the fact that string
field commutators do not vanish at spacelike separation, the
finiteness and interpretability of horizon entropy in string theory.
The task at hand now is to understand what all these phenomena imply
for a non-perturbative formulation of quantum gravity.

To begin, let us count the number of degrees of freedom of string
theory.  To do so, we need to make sure that any gauge symmetries of
the theory are completely fixed, since gauge symmetries are not
really symmetries at all, but redundancies in the description of the
theory.  The only such complete gauge fixing of string theory that we
know of is the light front gauge.

As we have seen, in light front gauge the longitudinal momentum of
the string is uniformly distributed over $\sigma$, and the
conservation of longitudinal momentum is equivalent to the
conservation of the total parameter length.  Let us regulate the
string theory by dividing the string into $N$ pieces, each with
longitudinal momentum $\varepsilon = p^+ / N$.  It is then true that
in any string interaction, the number of bits of string, or partons,
is conserved.

The free string Hamiltonian in light front gauge is then given by
\begin{equation}
H_0 = \frac{1}{2\varepsilon} \sum_i \left [ \Pv(i)^2 + \left (
\Xv(i+1) - \Xv(i) \right )^2 \right ] \>.
\end{equation}
The interaction Hamiltonian will include terms which generate the
usual splitting and joining of strings, as well as contact terms
which are necessary for gauge invariance.

There are some interesting aspects of the regulated light front
theory.  First of all, the full Hamiltonian depends only on the
transverse directions.  The interaction Hamiltonian contains only
delta-functions of transverse position.  There is no part of the
Hamiltonian which depends on the longitudinal direction.  On the face
of it, this fact seems very odd, and seems to indicate a severe
degree of non-locality in the theory.  We will soon see the
significance of this.

Second, if one adds more partons (increasing $N$ and simultaneously
decreasing the minimum longitudinal momentum $\varepsilon$) one finds
that the string does not develop short distance structure in
spacetime.  Instead, one obtains lots of very large loops, all with
radius of curvature of order $\sqrt{\alpha'}$ \cite{kks}.  These
loops are contained within an area of order $\log (N)$.  Thus, as we
increase the resolution, we do not obtain lots of small wiggles
around a nice, limiting curve, but instead find a wildly undulating
mess of string which grows very dense.

Finally, independent of string theory, suppose that the number of
degrees of freedom $N$ necessary to describe Planck scale physics
inside a volume $V$ was proportional to the volume.  This leads
immediately to a contradiction with black hole physics and the
Bekenstein-Hawking entropy \cite{holo}.  For suppose we have some
mass $M$ inside $V$.  If the boundary of $V$ has area $A$, drop in a
shall of mass that causes a black hole of area $A$ to form.  Then the
Bekenstein-Hawking entropy formula tells us that the number of
degrees of freedom necessary to describe Planck scale physics in the
volume {\sl after we have added some mass} is proportional to $A$.
Thus the entropy has {\sl decreased,} in direct contradiction to the
second law of thermodynamics.  Thus the maximum number of
non-redundant degrees of freedom necessary to describe Planck scale
physics inside a volume $V$ must in fact be proportional to the area
of the boundary.

Moreover, this suggests that, given a system inside a volume, there
exists a mapping of the system onto a set of surface degrees of
freedom which preserves the information contained within the original
system.  In particular, if we place a large screen outside a black
hole, there is a mapping from each Planck area on the event horizon
to a Planck area on the screen.

Are these ideas crazy?  It turns out that by using the focussing
theorem of general relativity, it is easy to show that, at least at
the classical level, one can in fact construct an injective map from
the surface of a black hole to the surface of a flat screen far away.
 Similarly, one can show that no matter how many black holes one puts
in a row, there always exists an injective map from the union of the
event horizons to the screen.  The conclusion, then, is that gravity
should be able to be formulated as a 2+1 dimensional theory with one
degree of freedom per Planck area \cite{holo}.

The significance of the strange non-locality we encountered in the
light front formulation of string theory is now becoming more clear.
As we saw above, the light front formulation of string theory, which
is fully gauge fixed and contains no redundant degrees of freedom,
does not depend at all on the longitudinal direction - it is a 2+1
dimensional theory.  Thus string theory is in some sense already a
holographic theory.

We saw earlier that as resolution time decreases (energy increases)
the string becomes more and more dense in spacetime.  When the
transverse density $\rho$ becomes of order $1/g^2$, interactions
become important, and we expect non-perturbative effects to enter.
This is, in fact, how the Planck scale enters string theory.  In
order for string theory to be consistent with the idea that there is
no more than one degree of freedom per Planck area, the
non-perturbative effects must be such that the string bits become
repulsive.  We can guess that the mean squared radius must start to
obey $R^2 \sim N$ beyond a certain momentum, presumably of order the
Planck energy.  This implies that we should expect a cross section
for Planck-scale scattering which grows like the energy at high
energy.

In conclusion, we see that string theory, if formulated as a
3+1-dimensional theory, must have so much gauge freedom that a
complete gauge fixing will reduce the theory to a discrete
2+1-dimensional theory.  This amount of gauge symmetry is far larger
than anything we have yet encountered.  We can easily conjecture that
the myriad dualities being discovered may be providing a first
glimpse at how this gauge symmetry is implemented in string theory.


\begin{thebibliography}{9}

\bibitem{aw}
J.~J.~Atick and E.~Witten,
{\sl Nucl. Phys.} {\bf B310} (1988) 291.

\bibitem{gm}
See, {\sl e.g.,} D.~J.~Gross and P.~F.~Mende,
{\sl Nucl. Phys.} {\bf B303} (1988) 407.

\bibitem{hawk}
S.~W.~Hawking,
{\sl Comm. Math. Phys.} {\bf 43} (1975) 199.

\bibitem{st}
L.~Susskind and L.~Thorlacius,
{\sl Phys. Rev.} {\bf D49} (1994) 966.

\bibitem{tpm}
K.~S.~Thorne, R.~H.~Price, and D.~A.~MacDonald,
{\sl Black Holes: The Membrane Paradigm,} Yale University Press,
1986.

\bibitem{bek}
J.~D.~Beckenstein,
{\sl Phys. Rev.} {\bf D7} (1973) 2333; {\sl Phys. Rev.} {\bf D9}
(1974) 3292.

\bibitem{hawktwo}
S.~W.~Hawking,
{\sl Phys. Rev.} {\bf D13} (1976) 191.

\bibitem{stu}
L.~Susskind, L.~Thorlacius, and J.~Uglum,
{\sl Phys. Rev.} {\bf D48} (1993) 3743.

\bibitem{dp}
D.~N.~Page,
{\sl Phys. Rev. Lett.} {\bf 71} (1993) 1291.

\bibitem{kks}
M.~Karliner, I.~Klebanov, and L.~Susskind,
{\sl Int. Jour. Mod. Phys.} {\bf A3} (1988) 1981.

\bibitem{sbhlc}
L.~Susskind,
{\sl Phys. Rev.} {\bf D49} (1994) 6606.

\bibitem{mpt}
A.~Mezhlumian, A.~Peet, and L.~Thorlacius,
{\sl Phys. Rev.} {\bf D50} (1994) 2725.

\bibitem{lsu}
D.~A.~Lowe, L.~Susskind, and J.~Uglum,
{\sl Phys. Lett.} {\bf B327} (1994) 226.

\bibitem{lpstu}
D.~A.~Lowe, J.~Polchinski, L.~Susskind, L.~Thorlacius, and J.~Uglum,
{\sl Black Hole Complementarity vs. Locality,} preprint
NSF-ITP-95-47, UCSBTH-95-12, SU-ITP-95-13, hep-th/9506138.

\bibitem{spec}
L.~Susskind,
{\sl Some Speculations About Black Hole Entropy in String Theory,}
Rutgers University preprint RU-93-44, August 1993, hep-th/9309145.

\bibitem{thooft}
G.~'t~Hooft,
{\sl Nucl. Phys.} {\bf B256} (1985) 727.

\bibitem{su}
L.~Susskind and J.~Uglum,
{\sl Phys. Rev.} {\bf D50} (1994) 2700.

\bibitem{verify}
J.-G.~Demers, R.~Lafrance, and R.~C.~Myers,
{\sl Phys. Rev.} {\bf D52} (1995) 2245; \hfil \break
D.~Kabat,
{\sl Nucl. Phys.} {\bf B453} (1995) 281.

\bibitem{kss}
D.~Kabat, S.~Shenker, and M.~J.~Strassler, {\sl Black Hole Entropy in
the $O(N)$ Model,} Rutgers University preprint RU-95-34,
hep-th/9506182.

\bibitem{holo}
L.~Susskind,
{\sl J. Math. Phys.} {\bf 36} (1995) 6377.

\end{thebibliography}
\end{document}